\documentclass[
reprint,
groupeaddress,
longbibliography,
bibnotes,
amsmath,amssymb,
aps,
prl,
]{revtex4-2}

\usepackage[utf8]{inputenc}
\usepackage[english]{babel}
\usepackage[plainpages = false, pdfpagelabels, 
                 bookmarks,
                 bookmarksopen = true,
                 bookmarksnumbered = true,
                 breaklinks = true,
                 linktocpage,
                 colorlinks = true,
                 linkcolor = blue,
                 urlcolor  = blue,
                 citecolor = blue,
                 anchorcolor = green,
                 hyperindex = true,
                 hyperfigures
                 ]{hyperref} 
\usepackage{enumerate} 
\usepackage{tabularx} 
\usepackage{array}    
\usepackage{tikz}
\usetikzlibrary{patterns}
\usepackage{slashed}
\usepackage{physics}
\usepackage{verbatim}
\usepackage{cancel}
\usepackage{bbold}
\usepackage{bm}
\usepackage{booktabs}
\usepackage[caption=false,singlelinecheck=off]{subfig}
\usepackage{blindtext}
\usepackage{bbm}
\usepackage{graphicx}
\usepackage{dcolumn}
\usepackage{bm}
\usepackage[normalem]{ulem}
\usepackage{mathtools}

\usepackage{pdfpages}
\usepackage{etoolbox} 

\makeatletter
\patchcmd{\@outputpage@head}{\@ifx{\LS@rot\@undefined}{}{\LS@rot}}{}{}{}
\makeatother

\DeclareRobustCommand{\svdots}{
  \vbox{%
    \baselineskip=0.33333\normalbaselineskip
    \lineskiplimit=0pt
\hbox{\scriptsize.}\hbox{\scriptsize.}\hbox{\scriptsize.}%
    \kern-0.2\baselineskip
  }%
}

\usepackage{xcolor}
\definecolor{MyViolet}{RGB}{153,  0,153}
\definecolor{FloBlue}{rgb}{0.25, 0.41, 0.88}

\begin{document}

\title{Giant Heat Flux Effect in Non-Chiral Transmission Lines}
\author{Florian St\"abler}
\affiliation{D\'epartement de Physique Th\'eorique, Universit\'e de Gen\`eve, CH-1211 Gen\`eve 4, Switzerland}
\author{ Alioune Gadiaga}
\affiliation{D\'epartement de Physique Th\'eorique, Universit\'e de Gen\`eve, CH-1211 Gen\`eve 4, Switzerland}
\author{Eugene V. Sukhorukov}
\affiliation{D\'epartement de Physique Th\'eorique, Universit\'e de Gen\`eve, CH-1211 Gen\`eve 4, Switzerland}

\begin{abstract}
We develop a theory of heat transport in non-chiral transmission lines (TLs) of quantum Hall edge channels coupled to Ohmic contacts (OCs) that accounts for a dynamical accumulation of charge in the reservoirs. As a consequence, heat transport is driven by charge fluctuations in the heat Coulomb blockade regime. This framework challenges conventional paradigms by revealing a giant heat flux effect—a significant amplification in heat transport arising from non-trivial fluctuation-dissipation relations. Through a Langevin-based approach, we derive the effective noise power in the chiral currents, which underlies this enhanced heat flux. Our findings predict clear experimental signatures unique to non-chiral TLs, as well as provide insights into finite-frequency effects, showing crossovers to more conventional diffusive behavior. This work offers a perspective on feedback mechanisms in quasi-1D heat transport with implications for dissipation control in low-dimensional quantum systems.
\end{abstract}

\date{\today}
\maketitle

\textit{Introduction}.---A key problem in Quantum Hall (QH) physics is understanding how disorder and interactions in QH edge states can disrupt charge and energy transport, potentially challenging the robustness of quantized electrical~\cite{Klitzing_qh} and thermal~\cite{Pendry1983Jul, Pekola21Oct, Kane1997Quantized} conductance. Recent experimental studies on mesoscopic QH circuits probe these properties by injecting and detecting charge and heat currents~\cite{le_sueur_energy_2010,granger_observation_2009,venkatachalam_local_2012,banerjee_observed_2017, Jezouin2013Nov}.

An essential component of mesoscopic QH circuits is the Ohmic contact (OC), a floating metallic granule coupled to 1D ballistic QH edge channels at various filling factors~\cite{Halperin1982Feb, Wen1990Jun, Buttiker1988Nov}. It serves as a probe for electrical and thermal properties~\cite{Buttiker1988Jan, Buttiker1988Nov, Battista2013Mar, Utsumi2014May, vandenBerg2015Jul, Dashti2018Aug}, while also introducing Coulomb interactions between the edge states, characterized by an energy scale determined by the granule's charging energy \(E_C\)~\cite{furusaki_theory_1995, Furusaki1998Mar, matveev_coulomb_1995, slobodeniuk_equilibration_2013}. At low temperatures where \(T/E_C \ll 1\), the Coulomb barrier inhibits charge and heat transport in classical systems~\cite{AverinCB} and acts as a band-pass filter in quantum systems. This leads to exciting phenomena such as dynamical renormalization of charge~\cite{duprez_dynamical_2021, altimiras_dynamical_2014} and heat transport~\cite{Sivre2018Feb, sivre_electronic_2019}, electron state teleportation~\cite{Idrisov2018Jul, Duprez2019}, Luttinger liquid behavior~\cite{anthore_circuit_2018, jezouin_tomonagaluttinger_2013}, tunable multi-channel Kondo effects~\cite{Iftikhar2015Oct, Iftikhar2018Jun}, charge fractionalization~\cite{idrisov_quantum_2020, morel_fractionalization_2022,Karki_double_2022,Karki_para_2023}, and Coulomb-mediated heat drag~\cite{idrisov_thermal_2022}.

A series of chirally connected Ohmic contacts (OCs) configured in a transmission line (TL) effectively modeled by floating, interacting reservoirs~\cite{stabler_transmissionline}, can be an effective model of disorder in QH edges due to QH puddle formation~\cite{shamim_counterpropagating_2022, hu2024resistancedistributiondecoherentquantum}. This setup enables theoretical studies that yield experimentally verifiable predictions, including models explaining how equilibrium is reached in disordered edge states~\cite{idrisov2023chargeconserving}, demonstrating that energy transport remains quantized despite dissipation, and offering insights into more exotic phenomena such as ``negative heat drag"~\cite{stabler_transmissionline, stabler_nonlocal}.

Originally proposed to resolve the “missing heat” paradox~\cite{goremykina_heat_2019, le_sueur_energy_2010, stabler_transmissionline, stabler_nonlocal}, the chiral TL framework focuses on the interacting regime where the OC charging energy is significant. Unexpectedly, it revealed unconventional physics, including a set of “magic” Lorenz numbers~\cite{Kiselev_wiedemann}, edge states that carry excess heat beyond the quantum limit~\cite{staebler_mesoscopic}, and behaviors closely linked to fractional statistics, resembling anyonic excitations~\cite{morel_fractionalization_2022}. In~\cite{staebler_mesoscopic}, we demonstrated that introducing a feedback loop breaks local chirality, enabling charge fluctuations to create correlated states through a backaction effect, causing all of these exotic phenomena.

\textit{In this Letter}, our motivation is two-fold: First, we extend the study of heat transport to globally non-chiral systems, a natural progression from previous work focused on chiral quantum Hall edge channels, expecting similar exotic physics. Specifically, we investigate a non-chiral TL composed of a series of OCs, as shown in Fig.~\ref{fig:setup}, under four boundary conditions: open, closed, periodic, and hybrid configurations (Fig.~\ref{fig:setup}(b-e)). Second, we contrast our approach with existing studies on non-chiral transmission lines, predominantly in the context of fractional QH states~\cite{park_tl_2019,spanslatt_conductance_2020,hein_thermal_2023,glattli2024toymodel23fractional,nosiglia_incoherent,Spanslatt_topological}. Unlike prior work, we explicitly incorporate the dynamical accumulation of charges \( Q_n(t) \) in reservoirs (\( \partial_t Q_n(t) \neq 0 \))---a critical aspect neglected in most models, which typically focus on zero-frequency correlation functions. However, we stress that in general, these effects cannot be neglected, as their relevance depends on the probed energy scales \( T/E_C \) and \( \tau_{\text{th}} T \), where the Thouless time \( \tau_{\text{th}} \)  will be discussed below.




Our setup offers another key advantage over other diffusive systems: the ability to individually probe the chiral current components in the TL. This can be achieved, for instance, through capacitive coupling or an attached quantum point contact, enabling precise exploration of fluctuation and response mechanisms.
\begin{figure}[t]
    \centering
    \includegraphics[width=1\linewidth,trim={0 .05cm 0 0},clip]{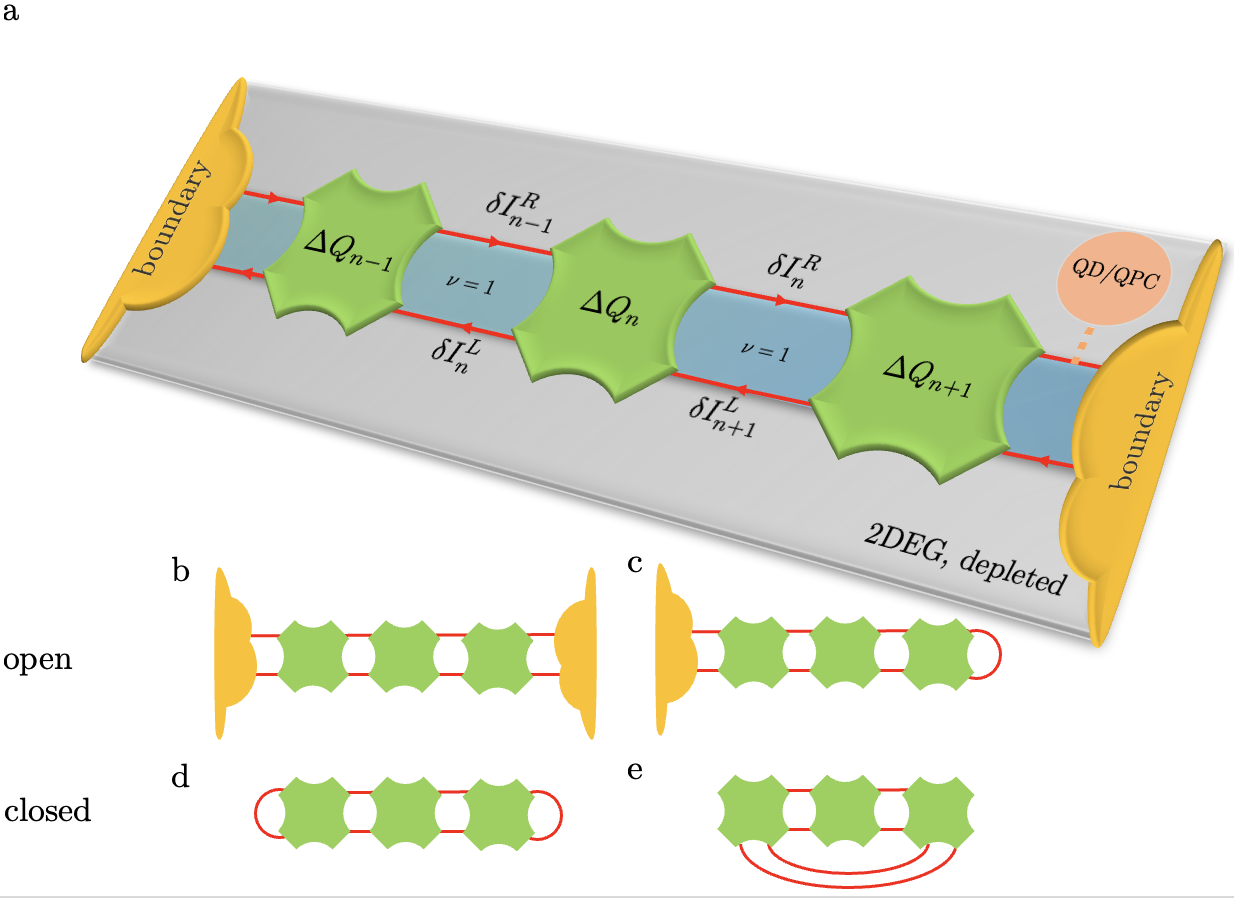}
   \caption{Schematic of a non-chiral transmission line (TL) comprising \( N = 3\) floating ohmic contacts (OCs) (green) connected to chiral one-dimensional quantum Hall edge channels (red arrows) at the boundary of a quantum Hall bulk with filling factor \(\nu = 1\) (light blue). The edge channels can extend from large metallic contacts (yellow) located at the left and right ends of the TL. The TL can be configured with various boundary conditions (b-e), including both open and closed system boundaries. Fluctuations in the edge channels induce charge fluctuations in the OC located at position \( n \) in the TL, denoted as \( \Delta Q_n \). These fluctuations generate charge current fluctuations \( \delta I^{\text{L/R}}_n \) in the outgoing channels to the left and right of each OC. For closed system boundary conditions (d and e), there is an additional constraint that total charge is conserved, meaning \( \sum_n \Delta Q_n(t) = 0 \) at all times. For illustration, panel (a) shows \( N = 3 \) OCs in the case of a fully open setup (b). The orange dot schematically indicates a measurement of the chiral current component, here \( \delta I_{n+1}^R \), using an adjacent quantum dot (QD) or quantum point contact (QPC).
}

    \label{fig:setup}
\end{figure}
Our key results are: (i) a nontrivial fluctuation-dissipation relation (FDR) in which noise power is proportional to the current response function of a specific current source, revealed by individual probing of chiral currents; (ii) a giant heat flux in open systems that greatly exceeds the equilibrium quantum of heat flux \( J_q = \frac{\pi}{12} T^2 \); and (iii) nontrivial electrical and thermal tunneling conductances that modify the Lorenz number, providing a distinctive signature for observing these altered edge states.

\textit{Solving the transmission line equation}.---We consider the TL schematically illustrated in Fig.~\ref{fig:setup}. The conservation of the OC charge \( Q_n(t) \) at position \( n \) leads to Kirchhoff's current law, relating the rate of change of \( Q_n(t) \) to incoming currents \( I^{\rm L}_{n+1}(t) \) and \( I^{\rm R}_{n-1}(t) \), and outgoing edge currents \( I^{\rm L}_{n}(t) \) and \( I^{\rm R}_{n}(t) \). This yields the TL equation:
\begin{align} 
    \frac{dQ_n(t)}{dt} &= I^{\rm L}_{n+1}(t)+ I^{\rm R}_{n-1}(t)- I^{\rm L}_{n}(t)- I^{\rm R}_{n}(t)\label{eq:Kirchhoff}.
\end{align} From now on we define the units such that $\hbar= k_B = e = 1$. The output charge current fluctuates due to two sources: the fluctuating OC electrical potential \( \Delta V_{n}(t) \equiv \Delta Q_n(t) / C \) with a finite capacitance \( C \), and neutral current fluctuations represented by the Langevin source \( \delta I^{\sigma,\text{c}}_n(t) \), arising from thermally induced uncertainty in the OC occupation:
\begin{align}
   \delta I^{ \sigma}_n(t) &= \frac{\Delta Q_n(t)}{\tau_C}  + \delta I^{ \sigma,\text{c}}_n(t), \quad \sigma = \text{L,R}\label{eq:Langevin} \ ,
\end{align} where \(\tau_C \equiv R_q C\) is the single-channel RC time constant, \(R_q \equiv 2\pi \) is the resistance quantum and \(C\) is the total capacitance. This time constant governs OC charge fluctuations and relates to the charging energy \(E_C \equiv 1/(2C) = \pi  / \tau_C\). 

In this section, we focus on the heat Coulomb blockade (HCB) regime, where the charging energy \(E_C\) significantly exceeds the OC temperature  \(T \ll E_C\), or equivalently, \(\tau_C T \ll 1\). By definition, fluctuations vanish on average: \( \langle \delta I^{\sigma,\text{c}}_n(t) \rangle = 0 \) and \( \langle \Delta Q_n(t) \rangle = 0 \). Inserting the Langevin equation~\eqref{eq:Langevin} into Kirchhoff's law~\eqref{eq:Kirchhoff} leads to a second-order inhomogeneous difference equation, which can be solved using a discrete Green's function \cite{Chung_Discrete, ellis2003discretegreensfunctionsproducts, Maximon_Difference}:
\begin{align} \label{eq:spectral_decomp}
    \Delta Q_n(\omega) = \sum_{m = 1}^N G_{n m}(\omega) s_m(\omega),
\end{align} where
$ s_m(\omega) \!=\! \delta I^{\rm L,c}_{m+1}(\omega)-
\delta I^{\rm L,c}_{m}(\omega) + \delta I^{\rm R,c}_{m-1}(\omega) - \delta I^{\rm R,c}_{m}(\omega)$. 
We can compute charge current fluctuations at each point \( n \) in the TL by substituting the formal solution~\eqref{eq:spectral_decomp} into the Langevin equation~\eqref{eq:Langevin}.

\textit{Discrete Green's Function techniques}.---The discrete Green's functions depend on the imposed boundary conditions (configurations b-e in Fig.~\ref{fig:setup}) and, in the HCB regime (\(\tau_C T \ll 1\)), satisfy the equation~\footnote{For closed boundary conditions (configurations d and e), the zero eigenvalue is excluded, resulting in a generalized Green's function as detailed in the supplemental material.}:
\begin{equation} \label{eq:gf_defining}
2 G_{nm} - G_{n+1m} - G_{n-1m} = \delta_{nm}.
\end{equation}
The Green's functions can be computed in three main ways. The first method relies on a spectral decomposition in terms of the eigenvalues \(\lambda_k\) and eigenvectors \(\phi_{n,k}\) of the finite difference operator \cite{Chung_Discrete,ellis2003discretegreensfunctionsproducts}:
\begin{align} \label{eq:decomp}
    G_{nm}(0) = \sum_{k = 1}^N \frac{1}{\lambda_k} \phi_{n,k} \phi_{m,k}.
\end{align}
A complete list of eigenvalues and eigenvectors for various boundary conditions is provided in the supplemental material.

Alternatively, the Green's function can be expressed as a product of two homogeneous solutions to the finite difference equation, similar to methods used for partial differential equations. As shown in Ref.~\cite{Maximon_Difference}, the Green's function takes the form
\begin{align}
    G_{nm}(0) \sim u_{1,n} u_{2,m},
\end{align}
where \( u_{p,n} \) satisfies the homogeneous equation \( 2 u_{p,n} - u_{p,n+1} - u_{p,n-1} = 0 \). The homogeneous solutions are constant or linear, indicating that \( G_{nm}(0) \) must be a constant, linear, or quadratic function of \( n \) and \( m \).

Finally, the Green's function can be directly computed by recursively applying the defining equation~\eqref{eq:gf_defining}, as detailed in the supplemental material. For example, under the open system boundary condition (configuration b in Fig.~\ref{fig:setup}), the Green's function is

\begin{equation}\label{eq:gf_open}
    G_{nm}(0) = \frac{n(N+1-m)}{N+1}, \quad n \leq m,
\end{equation}
with \( n \leftrightarrow m \) for \( n \geq m \). As expected, this result is a product of two linear functions in \( n \) and \( m \).

\textit{Fluctuation-Dissipation Relation of Chiral Currents in a Non-Chiral System}.---The formal solution of the equations of motion~\eqref{eq:Kirchhoff} and~\eqref{eq:Langevin}, expressed via the discrete Green's function~\eqref{eq:spectral_decomp}, allows us to compute current-current correlation functions in the transmission line: \(\langle \delta I^{\sigma}_n(\omega) \delta I^{\sigma'}_m(\omega') \rangle \equiv 2\pi \delta(\omega+\omega') S^{\sigma\sigma'}_{nm}(\omega)\), where \(S^{\sigma\sigma'}_{nm}(\omega)\) represents the noise power. Assuming fast relaxation in the OCs~\cite{spanslatt_impact}, the Langevin sources can be modeled by local thermal equilibrium noise power: \(\langle \delta I^{\sigma,c}_n(\omega) \delta I^{\sigma',c}_m(\omega') \rangle \equiv 2\pi \delta_{\sigma \sigma'} \delta_{nm} \delta(\omega+\omega') S^c_n(\omega)\),with source noise given by:
\begin{align}
  S^c_n(\omega)= \frac{\omega/R_q}{1 - e^{-\omega/T_n}}.  
\end{align}

For this letter, we assume uniform temperature across all OCs, \(T_n = T\), leading to \(S^c_n(\omega) = S^c(\omega)\), independent of \(n\). The non-equilibrium expression for the noise power was computed in the supplemental material. In equilibrium, \(S^{\sigma\sigma'}_{nm}(\omega)\) is given by:
\begin{multline}
    \label{eq:S_cross}
    S^{\sigma\sigma'}_{nm}(\omega) = S(\omega) \left( \delta_{nm} \delta_{\sigma\sigma'} \right. \\
    \left. + G_{nm+\text{s}(\sigma')}(\omega) + G_{mn+\text{s}(\sigma)}(-\omega) \right),
\end{multline}
where the sign of chirality is $\text{s}(L/R) = \pm 1$.

In the case of the autocorrelation function \( S^{\sigma\sigma}_{nn}(\omega) \) where the chirality and measurement position are identical, the noise power exhibits a modified Fluctuation-Dissipation Relation (FDR). This establishes a linear relationship between the current-current correlation function and the real part of the charge fluctuation response, as described by the Green's function:
\begin{align}
    \label{eq:S_auto}
    S^{\sigma\sigma}_{nn}(\omega) &= S^c(\omega) \left( 1 + 2\, \text{Re}\left[G_{n n+\text{s}(\sigma)}(\omega)\right] \right),
\end{align}

This result is significant because, in typical diffusive systems or Luttinger liquids, only non-chiral currents can be measured, corresponding to the net current across a cross-section, e.g., \(\delta I_n = \delta I_n^R - \delta I^L_{n+1}\). In contrast, chiral quantum Hall (QH) edge states enable direct probing of chiral current components, leading to a non-trivial response. Our key finding is that the equilibrium FDR for chiral currents in a non-chiral system is highly sensitive to the response function of charge fluctuations. We will discuss physical observables in which this unusual noise power \(S^{\sigma\sigma}_{nn}(\omega)\) plays a critical role, acting as a smoking gun for detecting the influence of finite OC charging energy.

\textit{Heat Flux and the Heat Coulomb Blockade Regime}.--- The heat flux \( J^\sigma_{n}(N) \) carried by the chiral edge state with chirality \(\sigma\) at a given position \(n\) in a TL of length $N$ is related to the noise power~\eqref{eq:S_auto} as follows~\cite{stabler_transmissionline}:
\begin{align}
\label{eq:T_integral}
    J^\sigma_{n}(N) = \frac{R_q}{4\pi} \int_{-\infty}^{\infty} d\omega \left[ S^{\sigma\sigma}_{nn}(\omega) - \lim_{T \rightarrow 0} S^{\sigma\sigma}_{nn}(\omega) \right],
\end{align}
where the noise power \( S^{\sigma\sigma}_{nn} \) is given by Eq.~\eqref{eq:S_auto}, and \( \lim_{T \rightarrow 0} S^{\sigma\sigma}_{nn}(\omega) \) represents the subtraction of vacuum fluctuations. In the heat Coulomb blockade regime, characterized by \( \tau_C T \ll 1 \), the heat flux can be evaluated as:
\begin{align}
\label{eq:hf}
    J^\sigma_{n}(N) = \left(1 + 2 G_{nn+\text{s}(\sigma)}(0)\right) J_q,
\end{align}
where \( J_q = \pi T^2 / 12 \) is the quantum of equilibrium heat flux. The heat flux depends on the boundary conditions imposed on the system, a summary of which can be found in Table~\ref{tab:res}.  For the TL in configuration b, the heat flux is given by
\begin{equation} \label{eq:hf_open}
    J^\sigma_n(N) / J_q = 1 + 2\frac{n(N-n)}{N+1},
\end{equation}
as shown in Fig.~\ref{fig:setup}. Most notably, for long TLs ($N \rightarrow \infty$) the heat flux is strongly enhanced leading to a what we call ``giant" heat flux effect.
\begin{figure}[ht!]
    \centering
        \includegraphics[width=\linewidth]{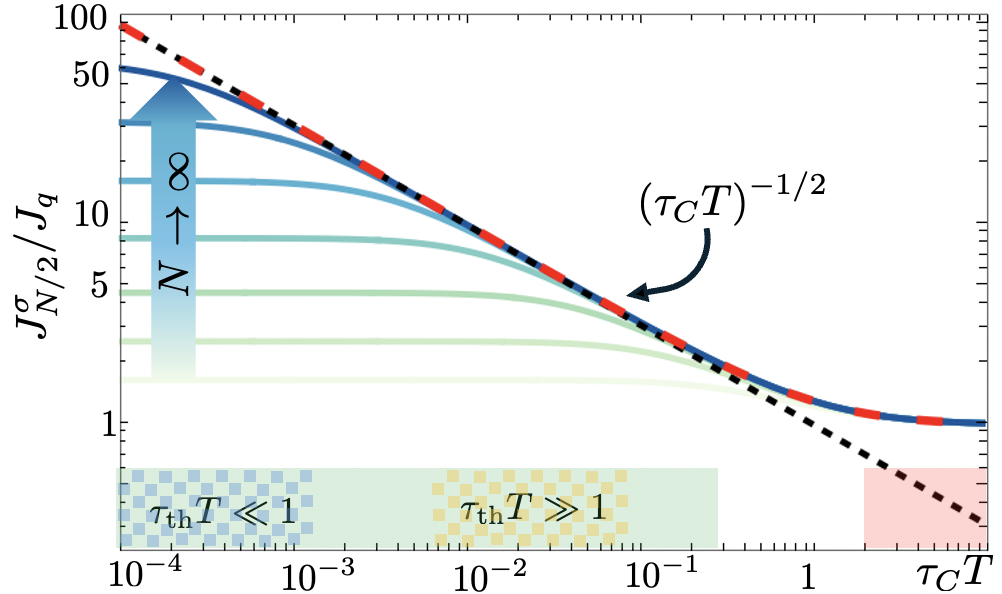}
\caption{Log-log plot of the heat flux at the TL center (\( n = \lfloor N/2 \rfloor \)) for open boundary conditions, as a function of the dimensionless product \( \tau_C T \). Results were obtained by numerically integrating the heat flux~\eqref{eq:T_integral} for different node counts \(N,\) accounting for the full frequency dependence of the Green's function.
In the high-temperature regime (\(\tau_C T \gg 1\), red region), the heat flux reaches the quantum heat flux limit.
In the heat Coulomb blockade regime (\(\tau_C T \ll 1\), green region), two distinct sub-regimes emerge based on the Thouless time \(\tau_{\text{th}}\): (i) For \( \tau_{\text{th}} T \ll 1 \) (blue-green region), we recover boundary-condition-sensitive flux values given by Eq.~\eqref{eq:hf_open}; (ii) For \( \tau_{\text{th}} T \gg 1 \) (orange-green region), a universal slope of  \(  J^\sigma_{N/2}/J_q \simeq 3 \zeta \left(\frac{3}{2}\right)/(\sqrt{2} \pi ^{3/2}) (\tau_C T)^{-1/2} \approx 0.9952(\tau_C T)^{-1/2} \) appears, to which the heat flux converges as \( N \to \infty \). The red dashed line represents the exact heat flux in the limit \( N \rightarrow \infty \). Explicit details of all the computations can be found in the supplemental material.}
    \label{fig:finite_freq}
\end{figure}

\textit{Finite frequency effect.}---To put the ``giant'' nature of this effect into perspective, the heat flux~\eqref{eq:hf_open} for open boundary conditions diverges linearly with system size for \( n \sim \lfloor N/2 \rfloor \). For a large number of nodes, the giant heat flux effect is ultimately constrained by the Thouless time \( \tau_{\text{th}} = N^2 \tau_C \), which sets the timescale where the finite-difference description of the system breaks down (e.g., for momenta above the inverse system size)
\footnote{For large \( N \), the discrete difference equation for \( Q_n \) approximates its continuous form as
\(
    \tau_C^{-1} \left(2Q_n - Q_{n+1} - Q_{n-1}\right) \approx \tau_C^{-1} \xi^2 \, \partial_x^2 Q(x),
\)
where \( \xi = L/N \) is the inter-node spacing. This approximation defines the Thouless time in terms of the effective diffusion constant \( D = \xi^2  \tau_C^{-1} \), yielding \( \tau_{\text{th}} \sim L^2 / D \sim N^2 \tau_C \).
}.

Our analysis so far assumes both the heat blockade regime, \( \tau_C T \ll 1 \), and the condition \( \tau_{\text{th}} T \ll 1 \), which imposes a stringent requirement, but allows us to neglect the frequency dependence of the response function. Beyond this regime, the finite frequency behavior shows a universal, boundary condition-independent  temperature dependence $\lim_{N\rightarrow \infty}J^\sigma_{N/2}/J_q =  3 \zeta \left(\frac{3}{2}\right)/(\sqrt{2} \pi ^{3/2})  \left(\tau_C T\right)^{-1/2}$, summarized in Fig.~\ref{fig:finite_freq}. Analogous arguments apply to the other boundary conditions listed in Table~\ref{tab:res}.
\renewcommand{\arraystretch}{3}
\begin{table*}[t]
\centering 
\begin{tabularx}{\linewidth}{p{2.5cm}p{4cm}p{3.5cm}p{5cm}p{2cm}}
\hline
\textbf{Observable}&\textbf{Open BC} & \textbf{Semi-Closed BC} & \textbf{Closed BC}& \textbf{Periodic BC} \\
\hline

$\displaystyle J^\sigma_n/J_q$& $\displaystyle 1+2\frac{n(N-n)}{N+1}$  & $\displaystyle 2n+ \sigma$ &  $\displaystyle \frac{2+(N-6) N+12\left(n-\frac{N}{2}\right)^2}{12 N}$&$\displaystyle \frac {5 + N^2} {6 N}$\\
$\displaystyle  L^\sigma_n(N)/L_0$ & $ \displaystyle \frac{3 \left(N+1+2 N n-2 n^2\right)}{3-2 n^2+N (3+2 n)}$ & $\displaystyle\frac{3 (2 n+\sigma )}{2+2 n+\sigma }$ &$\displaystyle \frac{3 \left(1+6 n^2-3 N-6 n N+2 N^2\right)}{1+6 n^2+9 N-6 n N+2 N^2}$&$\displaystyle\frac{3 \left(N^2+5\right)}{N (N+12)+5}$ \\
\end{tabularx}
\caption{Summary of findings in the heat Coulomb blockade regime $\tau_C T \ll 1$, in the low-temperature limit $\tau_{\text{th}}T \ll 1$, for different boundary conditions, including the heat flux $J^\sigma_n$ and Lorenz number $ L^\sigma_n(N)$, assuming the TL consists of $N$ OCs and current fluctuations are measured at position $n$. For the semi-closed boundary condition, it is assumed that the closed loop is to the right of the system $n=N$; a closed loop to the left can be computed by symmetry $n \leftrightarrow N-n$.}\label{tab:res}
\end{table*}

\textit{Electronic Correlation Function and Lorenz Numbers}.---The observation that charge fluctuations can lead to a non-quantized heat flux, \( J^\sigma_n(N) \neq J_q \), even in an equilibrium system, was discussed in~\cite{staebler_mesoscopic}. Let us consider an edge state subject to these additional charge fluctuations, connected to a quantum point contact (QPC). Measuring the electrical and thermal tunneling conductances allows us to extract a non-trivial, sometimes referred to as ``magic," Lorenz number~\cite{Kiselev_wiedemann,staebler_mesoscopic}, signifying a deviation from the Wiedemann-Franz law, which typically predicts a universal value \( L_0 = \pi^2 / 3 \) for the ratio of electrical to thermal conductance.

In a transmission line with \( N \) nodes, where current fluctuations are probed at position \( n \) with chirality \( \sigma \), the modified Wiedemann-Franz law is expressed as:
\begin{align}
\label{eq:WF_law}
    L^\sigma_n(N)/L_0 = 3 - \frac{6}{3 + 2 G_{nn+\text{s}(\sigma)}(0)},
\end{align}
where \( L_0 \) is the Lorenz number. This ratio reaches its maximum value (away from TL boundaries) of 3 in the regime \( \tau_{\text{th}} T \ll 1 \), with a crossover to 1 in the high-temperature regime \( \tau_C T \gg 1 \), even as \( N \rightarrow \infty \).

For the example of open system boundary conditions (configuration b), the Lorenz number is given by
\begin{align}
\label{eq:WF_law_open}
     L^\sigma_n(N)/L_0 =
    \frac{3 \left(N+1+2 N n-2 n^2\right)}{3-2 n^2+N (3+2 n)}
\end{align}
This result can be considered a smoking gun signature of correlated states. In~\cite{staebler_mesoscopic}, tunneling to the correlated states was assumed to be equivalent to tunneling directly to the OC metal, where the ``standard" \( P(E) \) theory applied. However, in the current scenario, non-zero cross-correlations, \( \left\langle \delta I^{\sigma,\text{c}}_n(t) \Delta Q_n(t) \right\rangle \neq 0 \), exist between the Langevin source and charge fluctuations. This suggests that the present findings are not fully captured by standard \( P(E) \) approaches, as the assumption of a free fermion moving in an electrical environment breaks down due to the presence of these additional cross-correlations. These correlations allow us to distinguish between tunneling to the chiral connecting edge state and tunneling directly to the OC, where such cross-correlations are absent.

\textit{Summary and Outlook}.---We have developed a theory of heat transport in non-chiral transmission lines formed by quantum Hall edge channels coupled to Ohmic contacts, where heat transport is significantly enhanced by charge fluctuations in the heat Coulomb blockade regime. This results in a ``giant heat flux effect"—a substantial amplification due to non-trivial fluctuation-dissipation relations. Using a Langevin approach, we derive effective (non-)equilibrium noise power of the chiral current components and give clear experimental signatures to unique non-chiral TL behaviors, providing a testable framework.

This work opens several avenues: (i) extending the model to fractional QH edge states, which could probe strongly interacting, disordered systems with diverse edge reconstructions; (ii) deriving non-equilibrium properties and examining modifications to Fourier’s law; (iii) applying this framework to 2D TL lattices to study bulk and edge effects in disordered 2D metals; (iv) exploring 2D random networks where Ohmic contacts may alter percolation physics and localization; and (v) examining TLs with finite-transparency QPCs, which would reveal more intricate Kondo physics compared to two site models currently investigated.

F.S. thanks M. Straub for fruitful discussions. The authors also thank C. Spånslätt, C. Glattli, and D. Karki for their useful comments on the manuscript. The authors acknowledges the financial support from the Swiss National Science Foundation.

\clearpage
\onecolumngrid

\begin{center}
    \textbf{\large Supplemental Material for ``Giant Heat Flux Effect in Non-Chiral Transmission Lines''}\\
    \vspace{0.5cm}
    Florian Stäbler, Alioune Gadiaga, and Eugene V. Sukhorukov\\
    \small\textit{Département de Physique Théorique, Université de Genève, CH-1211 Genève 4, Switzerland}\\
\end{center}

\title{Supplemental Material for ``Giant Heat Flux Effect in Non-Chiral Transmission Lines''}
\author{Florian St\"abler}
\affiliation{D\'epartement de Physique Th\'eorique, Universit\'e de Gen\`eve, CH-1211 Gen\`eve 4, Switzerland}
 \author{Alioune Gadiaga}
\affiliation{D\'epartement de Physique Th\'eorique, Universit\'e de Gen\`eve, CH-1211 Gen\`eve 4, Switzerland}
\author{Eugene V. Sukhorukov}
\affiliation{D\'epartement de Physique Th\'eorique, Universit\'e de Gen\`eve, CH-1211 Gen\`eve 4, Switzerland}
\date{\today}

\maketitle

\section*{Proof of the generalized FDT}
In this section, we prove the generalized Fluctuation Dissipation Theorem starting from the most general current-current correlation function and define the noise power, respectively.
\begin{subequations}
    \begin{align}
    \label{eq:S_def_cross_app}
    \langle  \delta I^{ \sigma}_n(\omega)  \delta I^{ \sigma'}_m(\omega') \rangle &\equiv 2\pi \delta(\omega+\omega')S^{\sigma\sigma'}_{nm}(\omega).
\end{align}
\end{subequations} Using the Langevin equation  and formal solution of the charge fluctuations in terms of the Green's function, we can evaluate the noise power $S^{\sigma\sigma'}_{nm}(\omega)$ as follows:
\begin{widetext}
\begin{multline} \label{eq:noise_non_eq}
    S_{nm}^{\sigma \sigma'}(\omega)   = (\delta _{nm} \delta _{\sigma \sigma'} +G_{mn+\text{s}(\sigma)}(-\omega)-G_{mn}(-\omega))S^{c,\sigma}_n(\omega)  +(G_{nm+\text{s}(\sigma')}(\omega)-G_{nm}(\omega)) S^{c,\sigma'}_m(\omega) \\  + \hspace{-.1cm} \sum
   _{n',\alpha = \text{L/R}}  \hspace{-.1cm} G_{nn'}(\omega) [G_{mn'}(-\omega) (S^{c,\alpha}_{n'-s(\alpha)}(\omega)+S^{c,\alpha}_{n'}(\omega))-G_{mn'+s(\alpha)}(-\omega) S^{c,\alpha}_{n'}(\omega)-G_{mn'-s(\alpha)}(-\omega)
   S^{c,\alpha}_{n'-s(\alpha)}(\omega)].
\end{multline}
\end{widetext} \onecolumngrid We assume that each OC is in local thermal equilibrium, i.e., the Langevin source correlation functions are given by $\langle  \delta I^{c, \sigma}_n(\omega)  \delta I^{c, \sigma'}_m(\omega') \rangle = 2\pi \delta_{nm} \delta_{\sigma\sigma' } \delta(\omega+\omega') S_n^{c,\sigma}(\omega)$, where $S_n^{c,\sigma}(\omega)$ is the equilibrium noise power $S_n^{c,\sigma}(\omega) = (\omega/R_q)\left(1-\exp(-\omega/T_n)\right)^{-1}$ that depends on the local equilibrium temperature $T_n$ of the OC at position $n$.  In a global equilibrium scenario, where all temperatures of the OCs are the same, i.e. $T_n = T$ and $S_n^{c,\sigma}(\omega) = S^c(\omega)$, the noise power~\eqref{eq:noise_non_eq} simplifies, using the defining equation of the Green's function Eq.(7) from the main text to
\begin{align} \label{eq:noise_eq}
   S^{\sigma\sigma}_{nn}(\omega)  &= S^c(\omega) \big(1 + 2 \text{Re}\left[G_{n  n+\text{s}(\sigma)}(\omega)\right] \big),\\
   S^{\sigma\sigma'}_{nm}(\omega) &=   S^c(\omega) \left( G_{nm+\text{s}(\sigma')}(\omega) +   G_{mn+\text{s}(\sigma)}(-\omega) \right).
\end{align} for the auto and cross noise power $(n \neq m)$, respectively. The equilibrium noise powers are linear in the Green's function, as expected for a Fluctuation-Dissipation relation. Note that the unsymmetrized noise power~\eqref{eq:noise_non_eq} is not necessarily real; however, when computing observables such as the heat flux, only the symmetric and thus real part of the noise power contributes.

\section*{Methods to Compute the Green's Function} 
Various methods exist to compute the finite difference Green's function. We will outline the different methods used throughout this paper, depending on which were the most feasible in each context. This appendix is based on approaches presented in \cite{Maximon_Difference,Chung_Discrete,ellis2003discretegreensfunctionsproducts}. Table~\ref{tab:gf} summarizes all Green's function subject to various boundary conditions.

\subsection*{Spectral Decomposition}
Using the Langevin and Kirchhoff equations from the main text, and considering the heat Coulomb blockade regime, we set $\tau_{C} \rightarrow 0$ and obtain the equation of motion for the charge fluctuations:
\begin{equation}
    2 \Delta Q_{n} - \Delta Q_{n+1} - \Delta Q_{n-1} = s_{n}(\omega),
\end{equation} where $s_n(\omega) = \delta I^{\rm L,c}_{n+1}(\omega) - \delta I^{\rm L,c}_{n}(\omega) + \delta I^{\rm R,c}_{n-1}(\omega) - \delta I^{\rm R,c}_{n}(\omega)$. Paralleling the Green's function method for   partial differential equations, this leads to a defining equation for the Green's function:
\begin{equation} \label{eq:gf_defining}
    2 G_{nm} - G_{n+1,m} - G_{n-1,m} = \delta_{nm}.
\end{equation}
The left-hand side can be viewed as the action of a linear operator $\mathcal{L}_{np} = 2\delta_{np} - \delta_{n+1,p} - \delta_{n-1,p} + \mathcal{B}_{np}$, subject to boundary conditions $\mathcal{B}_{np}$, acting on the Green's function such that $\mathcal{L}_{np} G_{pm} = \delta_{nm}$. For this operator, we can define eigenvalues $\lambda_k$ and eigenvectors $\phi_{n,k}$ through $\mathcal{L}_{np} \phi_{p,k} = \lambda_k \phi_{n,k}$. Note that $\mathcal{L}_{np}$ can be viewed as an $N \times N$ matrix. Any Green's function with $\lambda_k \neq 0$ may then be expressed as
\begin{align}\label{app:spec}
    G_{nm} = \sum_{k=1}^N \frac{1}{\lambda_k} \phi_{n,k} \phi_{m,k},
\end{align}
which can be verified by the action of $\mathcal{L}_{np}$ on $G_{pm}$.

\subsubsection*{Generalized Green's Function}
For closed and periodic system boundary conditions, the net charge in the TL is conserved, implying that the total change in charge across all sites is zero, i.e. $ \sum_n \Delta Q_n = 0$.  This conservation leads to one of the eigenvalues of the system being zero, $\lambda_{\Tilde{k}} = 0$, which results in an ill-defined Green's function. To address this, we define a modified Green's function by excluding the problematic eigenvalue:
\begin{equation} 
    G_{nm} = \sum_{k \ne \Tilde{k}} \frac{1}{\lambda_k} \phi_{n,k} \phi_{m,k}.
\end{equation}
This yields a revised equation for the modified Green's function:
\begin{align}
    \mathcal{L}_{np} G_{pm} = \delta_{nm} - \phi_{n,\Tilde{k}} \phi_{m,\Tilde{k}}^*,
\end{align}
where $\phi_{n,\Tilde{k}}$ represents the homogeneous solution to the equation $\mathcal{L}_{np} \phi_{n,\Tilde{k}} = 0$, associated with the zero eigenvalue. For both periodic and closed system boundary conditions, we have $\phi_{n,\Tilde{k}} = N^{-1/2}$, leading to the following equation for the Green's function:
\begin{align}
    \mathcal{L}_{np} G_{pm} = \delta_{nm} - \frac{1}{N}.
\end{align}

\subsection*{Explicit solution of  the Green's functions}
\subsubsection*{Open/Dirichlet Boundary conditions}

For the case of open system boundary conditions $\Delta Q_0 = \Delta Q_{N+1} = 0$, we can solve the recursion relation directly as follows. Let us assume we are in the low-temperature regime $\tau_{\text{th}} T \ll1$, for which the defining equation of the Green's function reads:
\begin{equation} \label{eq:app_defining}
2 G_{nm} - G_{n+1m} - G_{n-1m} = \delta_{nm}
\end{equation}
Considering the case $n<m$ and rearranging this equation gives:
\begin{equation} \label{eq_recurv}
G_{nm}-G_{n-1m} = G_{n-1m}-G_{n-2m}= \cdots = G_{1m},
\end{equation}
since due to the Dirichlet boundary condition $G_{0m} =0$. This means that the ansatz $G_{nm} = n G_{1m}$ solves Eq.~\eqref{eq_recurv} up to a constant factor. We can write another type of equation using the symmetry of $n$ and $m$:

\begin{equation} \label{eq:recurv_2}
     G_{1m}-G_{1m+1} = G_{1m+1}-G_{1m+2}= \cdots = -G_{1N},
\end{equation} which is solved by the ansatz $ G_{1m} = c (N+1-m)$ which allows us to write
\begin{equation}
G_{nm} = c n (N+1-m) \quad \text{for} \ n\leq m.
\end{equation} Due to the symmetry of the problem we have
\begin{equation}
G_{nm} = c m (N+1-n) \quad \text{for} \ n\geq m.
\end{equation} The coefficient $c$ can be determined from the equation
\begin{equation}
      2 G_{nm} - G_{n+1m} - G_{n-1m}  = 1, 
\end{equation}  which gives $ c  = \frac{1}{N+1}$, with the final solution
\begin{equation}\label{eq:gf0}
G_{nm} =  \frac{n (N+1-m)}{N+1} \quad \text{for} \ n\leq m,
\end{equation}  which allows to compute the nontrivial identity that follows immediately from the spectral decomposition~\eqref{app:spec} with $\lambda_k$ and $\phi_{n,k}$ found in Table \ref{tab:gf} for the open system case.
\begin{align}\label{eq:sumident}
G_{nm} = \frac{1}{N+1}\sum_{k=1}^N  \frac{\sin\left(\frac{kn\pi}{N+1}\right) \sin\left(\frac{k m\pi}{N+1}\right)}{1-\cos\left(\frac{k \pi}{N+1}\right)} 
=  \frac{n (N+1-m)}{N+1} \quad \text{for} \ n\leq m.
\end{align}

\subsubsection{Periodic Boundary Conditions}
For the case of periodic boundary conditions $\Delta Q_1 = \Delta Q_{N}$, we solve the recursion relation directly, following \cite{ellis2003discretegreensfunctionsproducts}. Let us assume we are in the heat Coulomb blockade regime $\tau_C T \ll1$, for which the defining equation of the Green's function reads:
\begin{align} \label{eq:eom_pbc}
    G_{a+1} - G_a = G_a - G_{a-1} + \frac{1}{N} - \delta_{a0},
\end{align} where $G_a \equiv G_{nm}$, with $ \ a \overset{!}{=} |m-n|$ indicates that the Green's function only depends on the difference of positions due to the periodicity of the system. The term $1/N$ appears because we consider the generalized Green's function as discussed above. Let us assume $a\neq0$ in equation of motion~\eqref{eq:eom_pbc} which yields after repeated application: 
\begin{align}\label{eq:recursion1_pbc}
    G_{a+1} - G_a = G_a - G_{a-1} + \frac{1}{N}  = G_{a-1} - G_{a-2} +   \frac{2}{N}
    =G_1 - G_0 + \frac{a}{N}  
    =\frac{1-N}{2N} +  \frac{a}{N},
\end{align} where the first term in the last equality can be obtained from the equation of motion~\eqref{eq:eom_pbc} setting $a =0$ and noting that  $G_{-1} \overset{!}{=} G_1$. Similar to before, we find the Green's function $G_a$ from the recursive application~\eqref{eq:recursion1_pbc} as:
\begin{multline}
    G_a = G_{a-1} + \frac{1-N}{2N}  + \frac{a-1}{N} 
    = G_{a-2} +  2 \frac{1-N}{2N} + \frac{a-1}{N} + \frac{a-2}{N} \\= G_0 + a \frac{1-N}{2N}  + \frac{1}{N} \sum_{k=1}^{a-1} k   = G_0 - \frac{a}{2} + \frac{a^2}{2N} 
\end{multline} 

Note that the TL with periodic boundary conditions is a closed system, which means that the total fluctuations of charge is conserved $\sum_n \Delta Q_n(t) = 0$ for all times $t$. This implies that the row sum of the Green's function is given by  $\sum_{a=0}^{N-1} G_a = 0$, from which we can compute $G_0 = (N+1)(N-1)/(12N)$. This yields the Green's function 
\begin{align}
    G_{nm} = \frac{N^2-1+6(m-n)^2}{12 N} - \frac{|m-n|}{2},
\end{align} and allows to compute the identity
\begin{align}
    G_{nm} = \frac{1}{N} \sum_{k=1}^N \frac{e^{i\frac{ 2 \pi }{N}  (m-n) k}}{4 \sin ^2\left(\frac{ \pi k }{N}\right)} 
    =\frac{N^2-1+6(m-n)^2}{12 N} - \frac{|m-n|}{2}.
\end{align}

\subsubsection{Closed Boundary Conditions}

For the case of closed system boundary conditions $\Delta Q_0 = \Delta Q_{1}$ and $\Delta Q_N = \Delta Q_{N+1}$, we solve the recursion relation directly as follows. Let us assume we are in the heat Coulomb blockade regime $\tau_C T \ll1$, for which the defining equation of the Green's function reads:
\begin{align} \label{eq:eom_closed}
    G_{nm+1} - G_{nm} = G_{nm} - G_{nm-1} + \frac{1}{N} - \delta_{nm},
\end{align}
For $n < m$, we obtain by successive lowering
\begin{align} \label{eq:lower}
      G_{nm+1} - G_{nm} = G_{nm} - G_{nm-1} + \frac{1}{N} = \cdots= G_{nn+1} - G_{nn} + \frac{m-n-1}{N}. 
\end{align} Next, we can use the equation of motion at $n = m$:
\begin{align}
    G_{nn+1} - G_{nn}  = G_{nn} - G_{nn-1} + \frac{1}{N}-1,
\end{align} which now allows us to lower the index $m$ further until we hit the left boundary condition by inserting in Eq.~\eqref{eq:lower} and successive lowering $n>m$ we obtain:
\begin{align} \label{eq:lower2}
      G_{nm+1} - G_{nm}  = G_{nn} - G_{nn-1} + \frac{1}{N}-1 + \frac{m-n-1}{N}  = \underbrace{G_{n1} - G_{n0}}_{= 0} +\frac{n}{N} + \frac{m-n-N}{N} = \frac{m-N}{N}.
\end{align}  Where the difference of Green's function vanishes due to the boundary condition $\Delta Q_0 = \Delta Q_1$. The identity for the difference of the Green's functions~\eqref{eq:lower2} allows us to write
\begin{multline}
    G_{nm} = G_{nm-1} + \frac{(m-1)-N}{N} =  G_{nm-2} + \frac{(m-2)-N}{N} + \frac{(m-1)-N}{N} = G_{nn} + \sum_{k=n}^{m-1} \frac{k}{N}-1 \\= G_{nn} + (m-n)  \biggl( \frac{m-1 +n }{2N} -1 \biggr),
\end{multline} where the only unknown $G_{nn}$ still needs to be determined. It can most easily be determined by solving the sum
\begin{align}
   G_{nn} = \frac{1}{2N} \sum_{k=1}^{N-1} \cot^2 \left( \frac{k\pi}{2N} \right) - \frac{2}{N} \sum_{k=1}^{N-1} \frac{\sin \left( \frac{k\pi}{N} (n-1) \right) \sin \left( \frac{k\pi}{N} n \right)}{4 \sin^2 \left( \frac{k\pi}{2N} \right)},
\end{align} which arises from the spectral decomposition of $G_{nm}$ and using the result that was obtained earlier for the Dirichlet Green's function~\eqref{eq:sumident} we obtain
\begin{align}
     G_{nn} = \frac{(2N-1)(N-1)}{6N} - \frac{(n-1)(N-n)}{N},
\end{align} and thus the full expression is
\begin{multline}
     G_{nm} = \frac{(2N-1)(N-1)}{6N} - \frac{(n-1)(N-n)}{N}  + (m-n)  \biggl( \frac{m-1 +n }{2N} -1 \biggr)\\=\frac{1+3 (m-1) m+3 (n-1) n+3 N-6 m N+2 N^2}{6 N},
\end{multline}  which yields the summation identity
\begin{align}
    G_{nm} = \frac{2}{N} \sum_{k=1}^{N} \frac{ \cos \left( \frac{\pi k}{N} (n - \frac{1}{2}) \right) \cos \left( \frac{\pi k}{N} (m - \frac{1}{2}) \right)}{4 \sin^2 \left( \frac{\pi k}{2 N}\right)} =\frac{1+3 (m-1) m+3 (n-1) n+3 N-6 m N+2 N^2}{6 N}.
\end{align}

\subsubsection{Semi-open Boundary Conditions}

For the semi-open boundary condition we revert to a different method and start directly from the spectral decomposition $ G_{nm}(0) = \sum_{k = 1}^N \lambda_k^{-1} \phi_{n,k} \phi_{m,k}$ with the eigenvalues and eigenvectors given by Table~\ref{tab:gf}:
\begin{align}
   G_{nm} =  \sum_{k=1}^{N} \frac{\sin\left(\frac{\pi n}{2N+1} (2k-1)\right) \sin\left(\frac{\pi m}{2N+1} (2k-1)\right)}{(2N+1) \sin^2 \left(\frac{\pi}{2} \frac{2k-1}{2N+1}\right)}. 
\end{align} 
We extend the summation to $2N+1$ by doubling the sum, which allows us to rewrite the Green's function in the following way:
\begin{align}
     G_{nm}= \sum_{k=1}^{2N+1}  \frac{1 + \cos{\theta_k}  }{2N+1} U_{n-1} (\cos{\theta_k})  U_{m-1} (\cos{\theta_k}) ,
\end{align} where $U_n(\cos{\theta_k}) = \sin \left((n+1)\pi \frac{2k-1}{2N+1}\right)/\sin \left(\pi \frac{2k-1}{2N+1}\right)$, with $\theta_k = \frac{2k-1}{2n+1}\pi$ are Chebyshev polynomials of the second kind. Using the identity \( U_{n+1}(\cos{\theta_k}) = 2 \cos{\theta_k} U_n(\cos{\theta_k}) - U_{n-1}(\cos{\theta_k}) \) and the orthogonality of Chebyshev polynomials of the second kind: 
\begin{align}
    \frac{1}{2N+1}\sum_{k=1}^{2N+1} U_n(\cos{\theta_k}) U_m(\cos{\theta_k}) = 
\begin{cases}
0 &  m - n \ \text{is odd}\\
 1 + \min(n,m) & m - n \ \text{is even}.
\end{cases}
\end{align} Note that  $1 + \min(n,m) = \min(n+1,m+1)$  and $\frac{1}{2}\bigl( \text{min}(n,m-1) + \text{min}(n,m+1)  \bigr) = \text{min}(n,m)$ leads to the simple form of the Green's function given by
\begin{align}
     G_{nm} = \text{min}(n,m), 
\end{align} and especially establishes the identity:
\begin{align}
   G_{nm} =  \sum_{k=1}^{N} \frac{\sin\left(\frac{\pi n}{2N+1} (2k-1)\right) \sin\left(\frac{\pi m}{2N+1} (2k-1)\right)}{(2N+1) \sin^2 \left(\frac{\pi}{2} \frac{2k-1}{2N+1}\right)} = \text{min}(n,m). 
\end{align}

\renewcommand{\arraystretch}{3}
\begin{table*}[htbp!]
\centering
\begin{tabularx}{\linewidth}{>{\raggedright\arraybackslash}p{.1\linewidth}>{\centering\arraybackslash}p{.2\linewidth}>{\centering\arraybackslash}p{.25\linewidth}>{\centering\arraybackslash}p{.45\linewidth}} 
\hline
\textbf{Boundary Condition} & \textbf{Eigenvalue} \( \lambda_k \) & \textbf{Eigenfunction} \( \phi_{n,k} \) & \textbf{Green's Function} \( G_{nm}(0) \) \\
\hline
\textbf{Open} & \(\displaystyle 4\sin^2\left(\frac{\pi k}{2(N+1)}\right)\) & \(\displaystyle \sqrt{\frac{2}{N+1}}\sin\left(\frac{n \pi k}{N+1}\right)\) & \(\displaystyle \frac{n(N+1-m)}{N+1}, \ n \leq m \) or  \( \displaystyle   \frac{m(N+1-n)}{N+1}, \ n \geq m \) \\
\textbf{Semi-Open} & \(\displaystyle 4\sin^2\left(\frac{\pi (k-\frac{1}{2})}{2N+1}\right)\) & \(\displaystyle \sqrt{\frac{2}{N+\frac{1}{2}}}\sin\left(\frac{n \pi (2k-1)}{2N+1}\right)\) & $\min\left(n,m\right)$ \\
\textbf{Closed} & \(\displaystyle 4 \sin^2\left(\frac{\pi (k-1)}{2N}\right)\) & \(\displaystyle \sqrt{\frac{2}{N}}\cos\left(\frac{(n-\frac{1}{2}) \pi (k-1)}{N}\right)\) & $\displaystyle\frac{1+3 (m-1) m+3 (n-1) n+3 N-6 m N+2 N^2}{6 N}$ \\
\textbf{Periodic} & \(\displaystyle 4\sin^2\left(\frac{\pi k}{N}\right)\) & \(\displaystyle \sqrt{\frac{1}{N}} \exp\left(i\frac{2\pi n k}{N}\right)\) & \(\displaystyle \frac{N^2 - 1 + 6(m-n)^2}{12N} - |m-n| \) \\
\end{tabularx}
\caption{Eigenvalues, Eigenfunctions, and Green's function \( G_{nm}(0) \) for various boundary conditions.}\label{tab:gf}
\end{table*}

\section{Green's Function at Finite Frequencies (Open System Boundary Conditions)}

We now consider open boundary conditions (configuration b in Fig. 1 in the main text) and extend the recursion approach to compute the Green's function at finite frequencies. The defining equation for the Green's function is
\begin{align}
(2 - i\omega \tau_C) G_{nm}(\omega) - G_{n+1m}(\omega) - G_{n-1m}(\omega) &= \delta_{nm}.
\end{align}
We introduce the mapping \( r + \frac{1}{r} = 2 - i \omega \tau_C \). For \( n \leq m \), this gives the recursion
\begin{align}
G_{n+1m}(\omega) - r G_{nm}(\omega) &= \frac{1}{r} \left( G_{nm}(\omega) - r G_{n-1m}(\omega) \right) = \frac{G_{1m}(\omega)}{r^n}.
\end{align}
For general \( n \leq m \), we obtain
\begin{align}
G_{nm}(\omega) &= \frac{G_{1m}(\omega)}{r^{n-1}} \frac{r^{2n} - 1}{r^2 - 1}.
\end{align}
Similarly, for \( n \geq m \), the recursion relation becomes
\begin{align}
G_{1m}(\omega) - r G_{1m-1}(\omega) &= r \left( G_{1m+1}(\omega) - r G_{1m}(\omega) \right) = -r^{N+2-m} G_{1N}(\omega),
\end{align}
which leads to the ansatz
\begin{align}
G_{1m}(\omega) &= c \left( r^2 - 1 \right) \left( 1 - r^{-2(N+1-m)} \right) r^{-1-m}.
\end{align}
Thus, for \( n \leq m \), the Green's function becomes
\begin{align}
G_{nm}(\omega) &= c \left( 1 - r^{-2(N+1-m)} \right) \left( r^{2n} - 1 \right) r^{-(n + m)}.
\end{align}
Due to symmetry, for \( n \geq m \), we have
\begin{align}
G_{nm}(\omega) &= c \left( 1 - r^{-2(N+1-n)} \right) \left( r^{2m} - 1 \right) r^{-(n + m)}.
\end{align}
To determine the coefficient \( c \), we use the condition
\begin{align}
(r + \frac{1}{r}) G_{nn}(\omega) - G_{n+1n}(\omega) - G_{n-1n}(\omega) &= 1,
\end{align}
which gives $c = r/((r^2 - 1)(1 - r^{-2(N+1)}))$. Finally, we obtain the Green's function for \( n \leq m \):
\begin{align}\label{eq:gf_ff}
G_{nm}(\omega) &= \frac{(r^n - r^{-n})(r^{N+1-m} - r^{-(N+1-m)})}{(r - r^{-1})(r^{N+1} - r^{-(N+1)})}.
\end{align}
In the heat Coulomb blockade regime (\( r \to 1 \)), the Green's function simplifies to Eq.~\eqref{eq:gf0}. The equilibrium noise power, as given by Eq.~\eqref{eq:noise_eq}, incorporates the Green's function~\eqref{eq:gf_ff}, where \( r \) satisfies a quadratic equation with two possible branches:
\begin{align}
r_{1/2} = 1 - i\frac{\omega \tau_C}{2} \pm \sqrt{-i \omega \tau_C - \frac{\omega^2 \tau_C^2}{4}},
\end{align}
as \( r \) is determined by the roots of this equation. However, after taking the real part of the Green's function, the result becomes independent of the branch choice. The heat integral, Eq.~(11) in the main text, then takes the form:
\begin{equation}\label{eq:heat_integral}
    J_n^{\sigma} - J_q = \int_{-\infty}^\infty \frac{d\omega}{2\pi} \, \text{Re}\left[\frac{(r^n - r^{-n})(r^{N-n} - r^{-(N-n)})}{(r - r^{-1})(r^{N+1} - r^{-(N+1)})}\right] \left(S^\sigma_n(\omega) - \omega \theta(\omega)\right),
\end{equation}
which produces the blue curves shown in Fig.~2 of the main text. In the bulk of the transmission line, at \( n = \lfloor N/2 \rfloor \), the Green's function~\eqref{eq:gf_ff} that enters the heat integral simplifies to:
\begin{align}\label{eq:gf_ff_simp}
    G_{\frac{N}{2},\frac{N}{2}+1}(\omega) \overset{N\rightarrow \infty}{=} \frac{r}{r^{-1} - r}.
\end{align}
Replacing this simplified Green's function into the heat integral~\eqref{eq:heat_integral} yields the red dashed line shown in Fig. 2 of the main text. To determine the prefactor of the asymptote for \( \tau_C T \rightarrow 0 \), we expand~\eqref{eq:gf_ff_simp}, yielding 
\( G_{\frac{N}{2},\frac{N}{2}+1}(\omega) \overset{N\rightarrow \infty}{\simeq} \left(2 |\omega| \tau_C T\right)^{-1/2} \), 
which results in the asymptotic heat flux:
\begin{equation}\label{eq:heat_integral_asymptote}
    J_n^{\sigma}/J_q \simeq \frac{1}{\sqrt{\tau_C T}} \int_{-\infty}^\infty \frac{d\omega}{4\pi} \frac{1}{\sqrt{2 |\omega|}}  
    \left(S^\sigma_n(\omega) - \omega \theta(\omega)\right) 
    = \frac{3 \zeta \left(\frac{3}{2}\right)}{\sqrt{2} \pi ^{3/2}} \frac{1}{\sqrt{\tau_C T}} ,
\end{equation}
where \( \zeta(x) \) is the Riemann zeta function and \(3 \zeta \left(\frac{3}{2}\right)/(\sqrt{2} \pi ^{3/2}) \approx 0.995215 \).







\clearpage
\twocolumngrid
\bibliography{References}

\end{document}